\documentclass[aps,showpacs,floats]{revtex4}
\usepackage{amssymb}
\usepackage{graphicx}
\usepackage{bm}

\begin{document}

\def\v#1{{\bf #1}}
\newcommand{\boldrho}{{\bm \rho}}
\newcommand{\la}{\langle}
\newcommand{\ra}{\rangle}
\newcommand{\lc}{\lowercase}

\newcommand{\hv}{\hat{v}}
\newcommand{\hvv}{$\hat{\v v}$}
\newcommand{\hvl}{$\hat{\v l}$}
\newcommand{\hvn}{$\hat{\v n}$}
\newcommand{\hvz}{$\hat{\v z}$}

\newcommand{\nn}{\nonumber}
\newcommand{\be}{\begin{eqnarray}}
\newcommand{\ee}{\end{eqnarray}}

\newcommand{\tphi}{\tilde{\varphi}}
\newcommand{\tthe}{\tilde{\vartheta}}
\newcommand{\trho}{\tilde{\rho}}
\newcommand{\tE}{\tilde{E}}
\newcommand{\tc}{\tilde{c}}
\newcommand{\tx}{\tilde{x}}
\newcommand{\ty}{\tilde{y}}
\newcommand{\tz}{\tilde{z}}
\newcommand{\tQ}{\tilde{Q}}

\newcommand{\De}{$\Delta$(\v k)}
\newcommand{\UPD}{UP\lc{d}$_2$A\lc{l}$_3$}

\title{Field-angle resolved specific heat and thermal conductivity in
the vortex phase of \UPD}
\author{P.~Thalmeier$^1$, T.~Watanabe$^2$, K.~Izawa$^2$ and Y.~Matsuda$^{2,3}$}
\affiliation{$^1$Max Planck Institute for Chemical Physics of
Solids, N\"othnitzer Str.40, 01187 Dresden, Germany\\
$^2$Institute for Solid State Physics, University of Tokyo,
Kashiwanoha 5-1-5, Kashiwa, Chiba 277-8581, Japan\\
$^3$Department of Physics, Kyoto University, Kyoto 606-8502, Japan}

\bibliographystyle{apsrev}

\begin{abstract}
The field-angle dependent specific heat and thermal conductivity in
the vortex phase of \UPD~is studied using the Doppler shift
approximation for the low energy quasiparticle excitations. 
We first give a concise presentation of the calculation procedure of
magnetothermal properties with vortex and FS averages performed
numerically. The comparison of calculated field-angle
oscillations and the experimental results obtained previously leads to
a strong reduction
of the possible SC candidate states in \UPD. The possible SC gap functions
have node lines in hexagonal symmetry planes containing either the zone
center or the AF zone boundary along c. Node lines in non-symmetry
planes can be excluded. We also calculate the field
and temperature dependence of field-angular oscillation amplitudes. We
show that the observed nonmonotonic field dependence and sign reversal
of the oscillation amplitude is due to small deviations from unitary scattering.
\end{abstract} 

\maketitle

\section{Introduction}
\label{sect:intro}

The U-based heavy fermion (HF) superconductors (SC) are supposed to
have unconventional SC order parameters usually (but not necessarily)
associated with anisotropic gap functions \De~that have node points or
lines on the Fermi surface (FS) where \De~= 0
\cite{Sigrist91,Thalmeier04}. This is thought to be the result of a
purely electronic mechanism of Cooper pair formation which favors
anisotropic gap functions due to a
strong on-site heavy quasiparticle repulsion. The type and position of
nodes in \De~is intimately connected with the symmetry class to which
\De~belongs, in most cases described by a single irreducible
representation of the high temperature symmetry group
\cite{Volovik85,Mineevbook}. Experimental evidence for the presence of
node lines is obtained from thermodynamic and transport
quantities as well as resonance experiments, but to locate their exact
position on the FS and hence
restrict the number of possible representations for the SC order
parameter is a difficult task. For example in UPt$_3$ it took a
considerable time
until it was identified as the two-dimensional odd
parity (spin-triplet) E$_{2u}$ representation and still there is no
unanimous agreement on this symmetry \cite{Joynt02}.

Recently the experimental determination of SC gap symmetries has been
much facilitated by the advent of a new method, namely the
investigation of field-angle dependence of specific heat and thermal
conductivity at temperatures T $\ll$ T$_c$. From the typical angular 
oscillations observed in these quantities under favorvable conditions
(small quasiparticle scattering) one may deduce the position of the
nodal lines or points of \De~with respect to the crystal
axis. Knowledge of these positions narrows down the possible choices
of representations for \De~considerably. This method has been
sucessfully applied to unconventional organic SC \cite{Izawa02}, to
ruthenates \cite{Izawa01a}, borocarbides \cite{Izawa02a,Park03} and Ce,Pr-based HF
superconductors \cite{Izawa01,Izawa03,Aoki04}. It is based on the 'Volovik effect'
\cite{Volovik93} which means the appearance of quasiparticle states in
the inter-vortex region of unconventional SC due to the presence of
nodes with \De~= 0 along certain directions in \v k-space. The zero energy
density of states (ZEDOS) of
these continuum states and hence their contribution to specific heat
and thermal conductivity depends on the relative orientation of field
direction, nodal positions and crystal axes through the
superfluid Doppler shift (DS) effect of quasiparticle energies. This
results in the angular oscillations of specific heat and thermal
conductivity which contain important information on the nodes of the
gap function. 

In addition this method has now been applied for the first time to a U-based
superconductor \cite{Watanabe04}, namely the intermetallic moderate
HF ($\gamma$ = 140 $\frac{mJ}{molK^2}$) compound
\UPD~\cite{Geibel91a}. This compound was
in the focus of interest in recent years because it is the only HF
superconductor where direct evidence for the microscopic nature of the
SC pairing mechanism has been found. This is connected with the fact
that \UPD~is the most clear cut example of a
U-based superconductor (SC) with dual-nature 5f electrons, some of
which are localised and some itinerant. The former can be considered as
5f$^2$ CEF states and the latter as conduction band
states \cite{Zwicknagl03}. The mass enhancement of conduction electrons
(m$^*$/m$_b\sim$ 10, m$_b$ is the band mass) is a result of
their interaction with the propagating CEF-excitations (`magnetic
excitons') associated with the localised 5f electrons. The
induced-moment AF order with
T$_N$ = 14.3 K, \v Q~= (0,0,0.5) (r.l.u.) and moderately large $\mu$ =
0.85$\mu_B$ coexists with SC
below T$_c$ = 1.8 K. In complementary INS \cite{Bernhoeft00,Sato01}
and quasiparticle tunneling experiments \cite{Jourdan99} both 5f
components were investigated and it was concluded \cite{Sato01} that
magnetic excitons mediate Cooper pairing. Theoretically this new
pairing mechanism was investigated in \cite{Thalmeier02,McHale04} and
possible symmetries of the SC states were discussed, also in
connection with existing Knight shift \cite{Kitaoka00} and upper
critical field results \cite{Hessert97}. The conventional itinerant
spin fluctuation mechanism has been investigated in
\cite{Nishikawa02,Oppeneer03} and also in \cite{Thalmeier02}.

The plausible SC gap functions obtained in \cite{McHale04} from a
microscopic model predict a node line parallel to the hexagonal
ab-plane but several D$_{6h}$ representations with different parity are
possible solutions. Furthermore the alternative
spin fluctuation model of \cite{Nishikawa02} predicts node lines
perpendicular to the basal plane. Therefore further investigation of
the gap structure of \UPD~has turned out to be necessary. It was
already suggested in \cite{Thalmeier02a} that field-angle resolved
experiments might be helpful to clarify the situation. They have now
indeed been performed in \cite{Watanabe04}.

The purpose of this paper is twofold: Firstly, although the theory of
magnetothermal properties in superconductors on the basis of the DS
approximation is well developed, the results are scattered through the
literature and therefore we first give a concise and complete outline
of the necessary computational steps for SC with uniaxial symmetry in
the superclean limit.
The calculation of linear specific heat coefficient $\gamma$(T,\v H) =
C(T,\v H)/T and thermal conductivity $\kappa_{ii}$(T,\v H) (i = x,z,y)
involves averaging over both
quasiparticle momenta and energies and the vortex coordinate. For
quantitative predictions the five-fold integrations are carried out
fully numerically for each of
the candidate gap functions. Also this has the advantage that one can
study the temperature dependence of oscillation amplitudes and real FS
geometry effects. Secondly we want to apply the DS theory in detail to 
\UPD~and study the predicted field-angle variations of the above
quantities for the possible gap functions with special
emphasis on the problem of node-line position along \v c$^*$. We also
discuss the influence of FS cylinder corrugation on field angle dependence
and the temperature variation of angular oscillation amplitudes and
investigate the dependence on the scattering phase shift.

In Sect.~\ref{sect:DSQP} we introduce the concept of the Doppler shift
approximation for quasiparticle energies and in Sect.~\ref{sect:SUPVEL} we give
the explicit expression for this quantity in two FS geometries. In
Sect.~\ref{sect:VORAV} we define the necessary averages over vortex coordinates
(superfluid velocity field) in the single-vortex approximation. The
calculation procedure for specific heat and thermal conductivity in
the superclean limit is given in Sect.~\ref{sect:THETRA}. Then in
Sect.~\ref{sect:UPDAL} we
apply the theory to \UPD~and discuss the results for the most
prominent candidate gap function in view of the available
experimental results in \cite{Watanabe04}. Finally
Sect.~\ref{sect:OUTLOOK} presents our conclusion on the gap symmetry
of \UPD~and an outlook on theoretical developments.

\section{Doppler shift of SC quasiparticles in the vortex phase}
\label{sect:DSQP}

In the vortex state the superfluid has acquired a velocity generated
by the gradient of the condensed phase as given by 
\be
\v v_s=\hbar\tensor{m}^{-1}(\nabla\phi-\frac{2e}{\hbar c}\v A) 
= \hbar\tensor{m}^{-1}\tilde{\nabla}\phi
\label{SUPVEL}
\ee
It is connected to the screening current circulating the vortex by 
\be
\v j_s(\v r)=2en_s(\v r)\v v_s(\v r)=\frac{c}{4\pi}\nabla
\times\v h(\v r)
\label{SUPCUR}
\ee
Here $m_{ij}=m_a\delta_{ij}+(m_c-m_a)\hat{n}_i\hat{n}_j$ is
the uniaxial mass tensor with $\hat{\v n}$ giving the
symmetry axis. Furthermore n$_s$(\v r) is the superfluid density and
\v h(\v r) is the local magnetic field strength. The above equations
hold for large Ginzburg parameters when local London
electrodynamics is applicable \cite{Balatskii86,Bulaevskii90}. 

Quasiparticle excitations out of the condensate with momentum \v p$_L$
have an energy E$_L$(\v p$_L$) in the local rest frame of the
superfluid. The transformation to the laboratory frame is given by the
universal law \cite{Volovikbook}
\be
\label{eq:DS}
\v p &=&\v p_L\nn\\
E(\v p,\v r)&=&E_L(\v p)+\v p\cdot\v v_s(\v r)
\label{EDS}
\ee
The second equation may be interpreted as a Doppler shift (DS) of
quasiparticle energies due to the moving condensate. In an
unconventional superconductor with gap nodes \De~= 0 on the FS
low energy (E $\ll\Delta$) quasiparticles can tunnel to the
intervortex region where they acquire a DS according to
Eq.~(\ref{eq:DS}). Since in a nodal SC the zero-field DOS starts like
a power law N(E) $\sim$ E$^n$ the DS, after averging over position \v
r and momentum \v p, will lead to a non-vanishing ZEDOS N(E=0,\v H) of
quasiparticles \cite{Volovik93} which determine the low temperature
(T $\ll$ T$_c$) specific heat and thermal transport.
%
\begin{figure}
\includegraphics[clip=true,width=70mm]{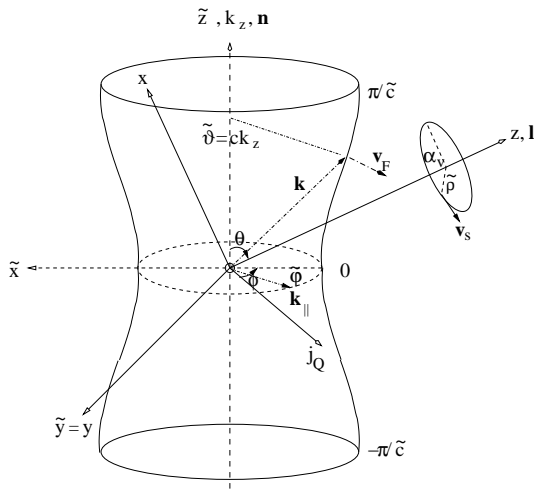}
\caption{Geometrical configuration of corrugated FS cylinder in the AF
BZ ($ -\frac{\pi}{\tc}\leq k_z\leq \frac{\pi}{\tc}$ with $\tc = 2c$),
vortex direction
(\v l) and quasiparticle momentum \v k. Here $\tx,\ty\equiv y, \tz$ is the
cartesian crystal coordinate system with a,b
and c (\v n)-axes. The cartesian coordinate system x,y,z is rotated by
the angle $\theta$ around y axis such that z ligns up with the vortex
direction \v l. Here $\phi$ is the angle between heat current \v j$_Q$ and the
field rotation ($\tilde{x}\tilde{z}$-) plane. Experimentally the perpendicular
configuration $\phi=90^\circ$ is used.
Note that for uniaxial crystals the vortex polar angle $\theta$
differs from the field polar angle $\theta_H$, this misallignment is
given by Eq.~(\ref{ANGTRANS}) and shown in the inset of
Fig.~\ref{fig:dopptot}. The quasiparticle momentum \v k is described
by cylindrical coordinates ($\tilde{\vartheta}=ck_z, \tilde{\varphi}$)
and \v v$_F$ $\perp$ FS denotes the quasiparticle velocity. The
superfluid velocity is denoted by $\v v_s$ and $\alpha_v$, $\trho$ are
the vortex coordinates.} 
\label{fig:geometry}
\end{figure}

\section{superfluid velocity in the London limit}
\label{sect:SUPVEL}

The superfluid velocity field \v v$_s$(\v r) is obtained from the field
distribution \v h(\v r) according to Eq.~(\ref{SUPCUR}). In the London
limit ($\lambda_i\gg\xi_i$; i=a,c with $\parallel\equiv$a, $\perp\equiv$c)
the latter is determined by the equation
\be
\v h +\lambda_\parallel^2\nabla\times(\nabla\times\v h)
+(\lambda_\perp^2-\lambda_\parallel^2)\nabla\times
[\v n(\v n\cdot\nabla\times\v h)] = \hat{\v z} \Phi_0\delta(\boldrho)
\label{LONDON}
\ee
where $\boldrho$=(x,y) are the cartesian coordinates of the plane
perpendicular to the vortex direction \hvl~$\parallel$ \hvz~
(Fig.~\ref{fig:geometry}) and $\Phi_0=hc/2e$ is the flux quantum.
In general \v h has components both parallel and
perpendicular to \hvl~\cite{Balatskii86}. Here we neglect the latter
and in the following assume \v h = h(x,y)\hvl. The superfluid velocity
and field
distribution in the xy-plane are not circular around the vortex in the
case of a uniaxial symmetry. However, we may apply a scale transformation
$\boldrho\rightarrow\boldrho'$ or $(x,y)\rightarrow$(x',y') given by 
\be
x'&=&\frac{x}{\lambda_\parallel}=\rho'\cos\alpha_v\\
y'&=&\frac{y}{\lambda_\theta}=\rho'\sin\alpha_v
\label{SPHERTRANS}
\ee
with $\rho'$=[x'$^2$+y'$^2$]$^\frac{1}{2}$ denoting the distance from
the vortex center, $\alpha_v$ the azimuthal angle around the vortex
and $\theta=\angle$(\hvl,\hvn), see Fig.~\ref{fig:geometry}. This
leads to a transformed h(x',y') determined by
\be
h(x',y')-(\partial_{x'}^2+\partial_{y'}^2)h(x',y')=
\frac{\Phi_0}{\lambda_\parallel\lambda_\theta}\delta(x',y')
\label{VORFIELD}
\ee
Therefore in the scaled x',y' coordinate system one has again a circular
vortex and field distribution. The solution of Eq.~(\ref{VORFIELD}) is
given by (K$_0$ = Hankel function) 
\be
\v h(\boldrho')=\frac{\Phi_0}{2\pi\lambda_\parallel\lambda_\theta}
K_0(\rho')\simeq
\frac{\Phi_0}{2\pi\lambda_\parallel\lambda_\theta}
[\ln\frac{1}{\rho'}+0.12]
\label{FIELDSOL}
\ee
The scaled superfluid velocity field 
v'$_x$=v$_x$/$\lambda_\parallel$, v'$_y$=v$_y$/$\lambda_\theta$ 
for distances $\rho'\gg\xi/\lambda$ is then obtained from Eq.~(\ref{SUPCUR}) as
\be
\v v'_s(\boldrho')=\frac{\hbar}{4m\lambda^2}\frac{1}{\rho'}
\left(\matrix{-\sin\alpha_v\cr\cos\alpha_v\cr 0}\right)
\label{VELSOL}
\ee
where $\lambda = (\lambda_\parallel\lambda_\theta)^\frac{1}{2}$,
$\xi = (\xi_\parallel\xi_\theta)^\frac{1}{2}$ and 
$m=(m_\parallel m_\theta)^\frac{1}{2}$ are defined in
App.~\ref{app:HCRIT}. The scaled $\rho'$ coordinate is dimensionless. We may
reintroduce length dimension but keep the circular vortex shape by
defining
\be
\trho =\lambda\rho'
\quad,\quad
\tilde{\v v}_s =\lambda\v v_s'
\quad,\quad
\label{SCALE}
\ee
In the rotated coordinate system (x,y,z) of Fig.~\ref{fig:geometry} with the
z-axis alligned with the vortex direction \hvl, the scaled quasiparticle
velocities defined in App.~\ref{app:QPVEL} are transformed to
\be
\v v_F'=\left(\matrix{
\lambda_\parallel^{-1}[v_a(\tthe)\cos\theta\cos\tphi +
v_c(\tthe)\sin\theta]\cr
\lambda_\theta^{-1}[v_a(\tthe)\sin\phi]\cr
\quad[-v_a(\tthe)\sin\theta\cos\tphi +v_c(\tthe)\cos\theta]}\right)
\label{QPVEL}
\ee
Here $v_i(\tthe)=v_i^0\hat{v}_i(\tthe)$ where $\tthe$ is the polar angle for an 
ellipsoidal FS and $\tthe =ck_z$ for the cylindrical FS of
App.~\ref{app:QPVEL}. Furthermore ($\tphi$,$\tthe$) defines the direction of
quasiparticle momentum
on the FS and $\theta$ is the vortex tilt angle in the field rotation
plane which for convenience is chosen as the $\tilde{x}\tilde{z}$-plane.

The dimensionless Doppler shift energy x (not to be confused with the
cartesian coordinate) for a given vortex
direction is then given by
\be
x=\frac{\v p_F\cdot\tilde{\v v}_s}{\Delta}=
\frac{1}{\Delta}m\tilde{\v v}_F(\tthe,\tphi;\theta)
\cdot\tilde{\v v}_s(\trho,\alpha_v)
\label{DOPPDEF}
\ee
with $\tilde{\v v}_F=\lambda\v v_F'$ and  $\tilde{\v v}_s$ defined by
Eqs.~(\ref{VELSOL}),(\ref{SCALE}) and $\Delta$ denoting the SC gap
amplitude. Using Eqs.~(\ref{VELSOL}),(\ref{QPVEL}) we finally obtain for x =
x($\theta,\tthe,\tphi;\trho,\alpha_v$) the explicit result
\be
x=\frac{1}{4}\frac{\xi_a^0}{\trho}\Bigl\{
\bigl(\frac{\lambda_\parallel}{\lambda}\bigr)
\hv_a(\tthe)\sin\tphi\cos\alpha_v-
\bigl(\frac{\lambda_\theta}{\lambda}\bigr)
[\hv_a(\tthe)\cos\theta\cos\tphi +
\alpha^\frac{1}{2}\hv_c(\tthe)\sin\theta]
\sin\alpha_v\Bigr\}
\label{DOPPEXP}
\ee
where $\sqrt{\alpha}=v_c^0/v_a^0$ is the anisotropy of the Fermi
velocity, $\xi_a^0=(\hbar v_a^0/\pi\Delta)$ is the in-plane
coherence length and $\hv_{a,c}(\tthe)$ are given in
App.~\ref{app:QPVEL}. Thus the Doppler shift energy of a quasiparticle
depends on three sets of variables: The vortex direction
($\theta$) (or field direction $\theta_H$), see
App.~\ref{app:HCRIT}),
the quasiparticle momentum coordinates ($\tthe,\tphi$) on the FS and
the quasiparticle position in real space ($\trho,\alpha_v$) with
respect to the vortex center. The expression in Eq.(\ref{DOPPEXP}) is
the central quantity which determines the thermodynamic and transport
properties in the vortex phase. In the special case for \v H
$\parallel$ c ($\theta=0$) the DS simplifies to 
\begin{equation}
x=\frac{1}{4}\frac{\xi_a^0}{\trho}\hv_a(\tthe)\sin(\tphi-\alpha_v)
\label{DOPPEXP0}
\end{equation}
Because x now depends only on the angle difference $\tphi-\alpha_v$,
averaging over the DS for \v H $\parallel$ c involves one
integration less than for general field direction.

\section{Average over the vortex coordinates}
\label{sect:VORAV}

The calculation of thermodynamic and transport coefficients and the
quasiparticle DOS involves averaging over both vortex coordinates
(real space) and quasiparticle velocities (momentum space). It is
instructive to calculate the vortex averaged DS energy 
$\la |x|\ra^{\v k}_v$ of quasiparticles with given \v k,
i.e. ($\tthe,\tphi$), as function field direction $\theta$. Since the important
contributions to $\gamma$(T,\v H) and $\kappa_{ii}$(T,\v H) come from
the nodal regions with $|\Delta(\v k)|/\Delta\leq|x|$, the field-angle
dependence of $\la |x|\ra^{\v k}_v$ for \v k $\in$ nodal region gives
already qualitative information on the behaviour of $\gamma$ and
$\kappa_{ii}$ as shown in this section. 

In the scaled coordinates x',y' spanning the plane perpendicular to
the vortex direction \hvl($\theta$) the field distribution h(x',y')
and velocity distribution \v v$_s$(x',y') are circularly symmetric for
a single vortex. In the limit H $\ll$ H$_{c2}$ 
the average over the
vortex lattice is then approximately replaced by the integration over a
single rotationally symmetric vortex according to
\be
\la A(\trho,\alpha_v)\ra_v=\frac{1}{\pi(\tilde{d}^2-\tilde{d}_c^2)}
\int_0^{2\pi}d\alpha_v\int_{\tilde{d}_c}^{\tilde{d}}d\trho\trho
A(\trho,\alpha_v)
\label{VORAV}
\ee
Here $\tilde{d}_c$ is a lower cutoff of the size of the coherence
length $\xi=(\xi_\parallel\xi_\theta)^\frac{1}{2}$ and $\tilde{d}$ is
of the order of the inter-vortex distance. It is determined by
assuming a square vortex lattice and replacing its (square) unit cell
of area F$_\Box$ by a circle of equal area with radius $\tilde{d}$
requiring $\pi\tilde{d}^2$ = F$_\Box$=$\Phi_0$/H. This leads to
\be
\tilde{d}=\tilde{d}_0\bigl(\frac{H}{H_0}\bigr)^{-\frac{1}{2}}
\quad;\quad
\tilde{d}_0=\bigl(\frac{\Phi_0}{\pi}\bigr)^\frac{1}{2}\frac{1}{\sqrt{H_0}}=
\bigl(\frac{hc}{2\pi}\bigr)^\frac{1}{2}\frac{1}{\sqrt{eH_0}}
\label{MAGSCL}
\ee
%
\begin{figure}
\includegraphics[clip=true,width=70mm]{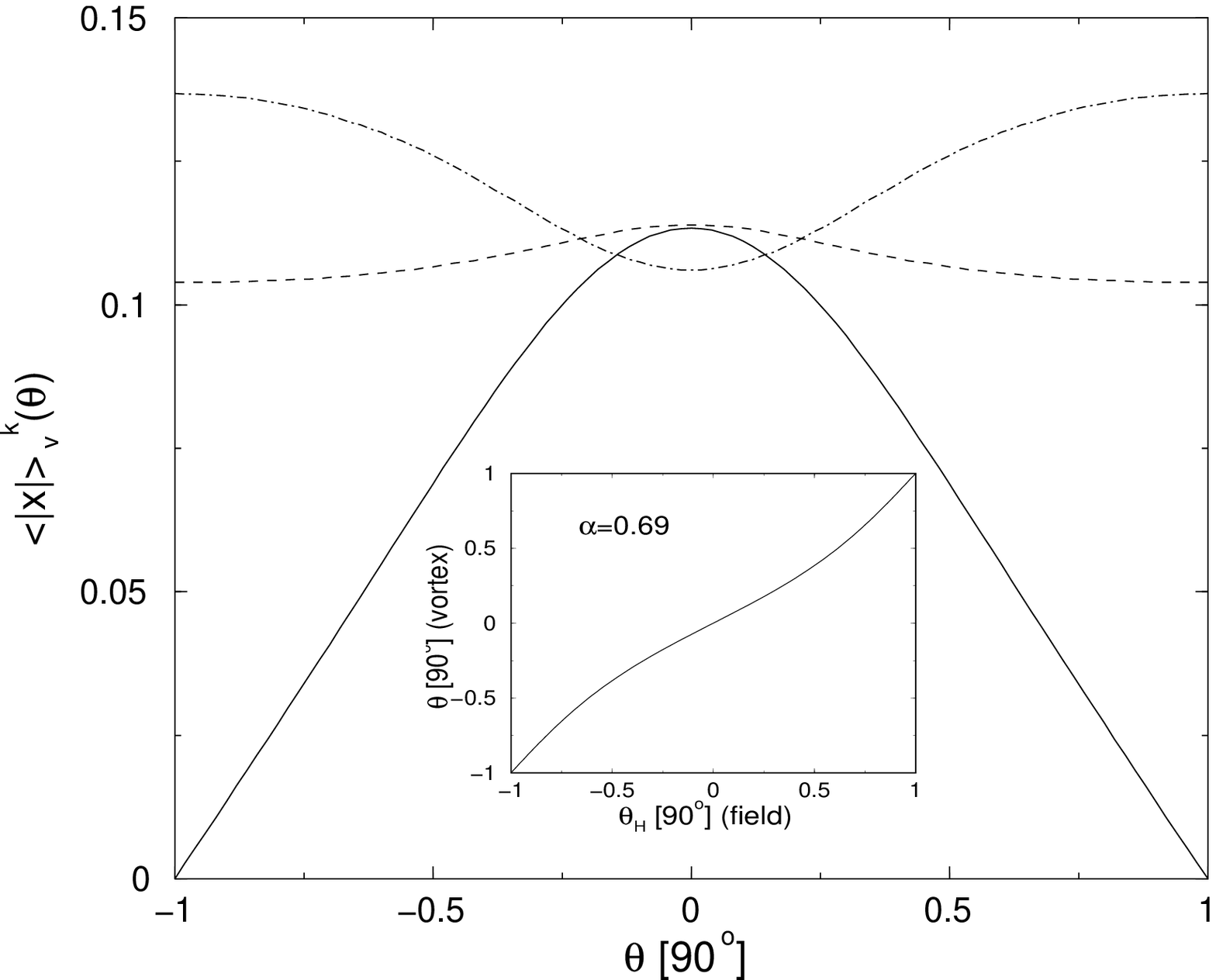}
\caption{Vortex-averaged dimensionless Doppler shift (H = 1T) as function of
vortex direction $\theta$ for various quasiparticle momenta
\v k which
are given in cylindrical coordinates as ($\tthe=c
k_z,\tphi$). Full line: ($\frac{\pi}{2}$,0); dashed line:
($\frac{\pi}{2}$,$\frac{\pi}{2}$); these \v k-vectors lie in the
k$_z$=$\pi/2c$ symmetry plane which is also the nodal AF zone-boundary
plane for A$_{1g}$.  dash-dotted line:
($\frac{\pi}{4}$,$\frac{\pi}{2}$); this \v k-vector lies in a
non-symmetry plane which 
is the nodal plane for the last gap function in
Table~\ref{tab:GAPFUN}. The small splitting at $\theta = 90^\circ$ is
caused by the FS corrugation. The inset shows the misallignment between
vortex ($\theta$) and field ($\theta_H$) directions for a mass
anisotropy $\alpha$=0.69 appropriate for \UPD. In all following
angular plots we use $\theta$ as variable and neglect the difference
between $\theta$ and $\theta_H$.}
\label{fig:dopptot}
\end{figure}
%
with $\Phi_0$=(hc/2e) = 2.07$\cdot$10$^{-11}$Tcm$^2$ and H$_0$ = 1T the
magnetic length scale involved is $\tilde{d_0}$(1T) = 257 \AA. This
means $\tilde{d}_0/\xi_a^0=3$ when we approximate the BCS or Pippard
coherence length $\xi_a^0$ by the
a,c- averaged value $\xi_0\sim$ 85\AA~given in
Ref.~\cite{Geibel91a} for \UPD. This estimate has some uncertainty
depending on the inclusion of Pauli limiting effects. The ratio
$\tilde{d}_0/\xi_a^0$ determines
directly the size of the DS via the prefactor in Eq.~(\ref{DOPPEXP0})
when $\trho$ is expressed in units of $\tilde{d}_0$. 
Decreasing the ratio $\tilde{d}_0/\xi_a^0$ at fixed H($\tilde{d}_0$)
increases the DS and hence the oscillation amplitudes. It is useful to
check the consistency of this value in an independent way. Using the
expression for H$_{c2}$ in Eq.~(\ref{HCRIT}) we may also write the
magnetic length scale as
\be
\tilde{d}_0=\sqrt{2}\xi_a^0\Bigl(\frac{H_{c2}}{H_0}\Bigr)^\frac{1}{2}
\ee
With the experimental H$^a_{c2}$ = 3.2T \cite{Watanabe04} we obtain
$\tilde{d_0}$(1T) = 215 \AA~which is consistent with the previous
value. The discrepancy may be due to Pauli limiting effects which
reduce H$_{c2}$ \cite{Hessert97} from its purely orbital value. These
estimates also give an insight in the
validity range of the single vortex and DS approximation. 
For fields H $\ll$ H$_{c2}$ (but large enough to fulfill $|x|\Delta\gg\
\Gamma$) we have
$\xi_0\sim\tilde{d}_c\ll\tilde{d}$, consequently we may approximate
$\tilde{d}_c\simeq 0$ in this limit. This means we commit a small
error by extending the DS approximation to the core region where it is
not valid.\\ 

We now perform the average in Eq.~(\ref{VORAV}) for
$A(\trho,\alpha_v)=|x(\theta;\tthe,\tphi;\trho,\alpha_v)|$,
i.e. the absolute value for the DS energy in Eq.~(\ref{DOPPEXP}) for fixed
momentum \v k ($\tthe,\tphi$) direction. In Fig.~\ref{fig:dopptot} we compare
$\la|x|\ra^{\v k}_v$ for various quasiparticle momenta
\v k. Keeping in mind that according
to Eq.~(\ref{DOPPDEF}) $|x|$ vanishes when $\tilde{\v v}_F\perp\tilde{\v v}_s$
and becomes maximal for $\tilde{\v v}_F\parallel\tilde{\v v}_s$, the
$\theta$ (or $\theta_H$) variation of $\la|x|\ra^{\v k}_v$ in
Fig.~\ref{fig:dopptot}
can be qualitatively understood. It is the $\theta$ variation
of $\la|x|\ra^{\v k}_v$ for \v k $\in$ nodal region which survives in 
the ZEDOS and transport coefficients from which conclusions on
the positions of nodes of \De~may be drawn.

\section{Thermodynamics and transport in the vortex phase in DS
approximation}
\label{sect:THETRA}

The low temperature transport and thermodynamics in unconventional SC
are determined by the combined effect of impurity scattering and DS
due to the supercurrents. Both effects may lead to a low energy
residual DOS which influences specific heat and thermal conductivity. In
the zero-field limit the theory is well developed
\cite{Mineevbook}. Our intention here is to study typical signatures
of the nodes of \De~in magnetothermal properties to draw conclusions
on the gap structure. For this purpose the 'superclean limit' where
the DS energy dominates the effect of scattering ($|x|\Delta\gg\Gamma$)
is the relevant one. 

Transport and thermodynamics of unconventinal SC in DS approximation
has been developed by many authors over the years, following the
pioneering work of Volovik \cite{Volovik93}, we mention only a few of
them here \cite{Barash97,Kuebert98,Vekhter01,Won00,Dahm00,Won01a,Won01}.
For our purpose it is sufficient to have a summary of these results in
the superclean limit in a concise form useful for numerical
computation of the field-angle dependence of $\kappa_{ii}(\v H,T)$ and
$\gamma(\v H,T)$. 
%
\begin{figure}
\includegraphics[clip=true,width=70mm]{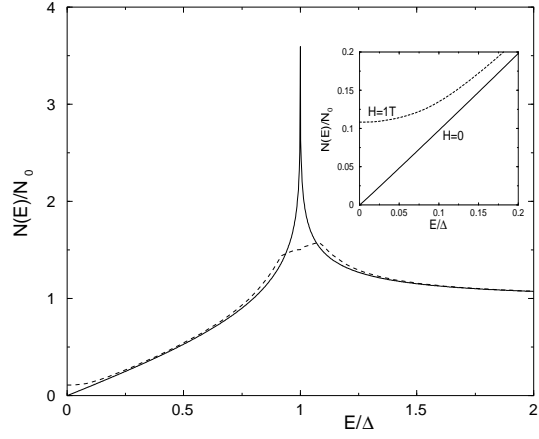}
\caption{Quasiparticle DOS for the A$_{1g}$ gap function for
H = 0 (full line) and 1T (dash-dotted line)  and field $\parallel$ c
($\theta_H = \theta =0$).
 The DOS singularity is smeared out by the
\v k-dependent DS. Inset shows the ZEDOS induced by the Doppler
shift on an enlarged scale.}
\label{fig:dos}
\end{figure}
%
In the zero-field case the quasiparticle energies of an unconventional
superconductor are given by E$_{\v k}$ = [$\epsilon_{\v k}^2$+
$\Delta_{\v k}^2$]$^\frac{1}{2}$ where $\Delta_{\v k}$ is the
nontrivial singlet gap
function or $\Delta_{\v k}^2= |\v d_{\v k}|^2$ with $\v d_{\v k}$
denoting the (unitary) triplet gap functions. The \v k ($\tthe,\tphi$)
dependence of the gap function may be characterised by the form factor
$\Phi$(\v k)=\De/$\Delta$ where $\Delta$(T) is the gap amplitude obtained
from the solution of the gap equation.

In the vortex phase the DS leads to an additional position
($\trho$,$\alpha_v$) and field-angle ($\theta$) dependence of
quasiparticle energies according to Eq.~(\ref{DOPPEXP}). Defining E'$_{\v k}$
= E$_{\v k}$/$\Delta$ we obtain for \v k ($\tthe,\tphi$):
\be
\tE_{\v k}(\theta;\trho,\alpha_v)=
E'_{\v k}-x_{\v k}(\theta;\trho,\alpha_v)
\label{ETIL}
\ee
Calculations of $\gamma(\theta)$ and $\kappa_{ii}(\theta)$
therefore involves, in addition to the FS averaging present already in
the zero-field case, the averaging over vortex coordinates as
prescribed in Eq.~(\ref{VORAV}).

\subsection{Quasiparticle DOS and specific heat}
\label{subsect:QPDOS}

In the zero-field case ($\Gamma\rightarrow 0$) the quasiparticle DOS
is given by 
\be
N(E)/N_0&=&g_1(E) \quad\mbox{with}\quad g_1(E) = Re~g(E)\nn\\
g(E)&=&g_1(E)+ig_2(E)=\int dS_{\v k}
\frac{E}{[E^2-\Delta_{\v k}^2]^\frac{1}{2}}
\label{DOS}
\ee
where dS$_{\v k}$ = $\frac{1}{4\pi}$d$\Omega_{\v k}$= $\frac{1}{4\pi}\sin\tthe
d\tthe d\tphi$ or dS$_{\v k}$ =$\frac{\tilde{c}}{4\pi^2}d\tphi dk_z$
for the ellipsoidal and cylindrical FS case respectively. 
In the vortex phase one has to replace E by the Doppler shifted
quasiparticle energies and form averages $G_{1,2}(E)=\la
g_{1,2}(E)\ra_v$ over the vortex coordinates according the
prescription of Eq.~(\ref{VORAV}). Then G$_{1,2}(E)$ and the field-angle
dependent quasiparticle DOS are given explicitly as
\be
N(E,H,\theta)/N_0 = G_1(E) &=& \Bigl\la\int dS_{\v k}
\frac{\frac{1}{\sqrt{2}}|\tE_{\v k}|(|\tE_{\v k}^2-\Delta_{\v k}^2|
+\tE_{\v k}^2-\Delta_{\v k}^2)^\frac{1}{2}}
{|\tE_{\v k}^2-\Delta_{\v k}^2|}\Bigr\ra_v \nonumber\\
G_2(E) &=& -\Bigl\la\int dS_{\v k}
\frac{\frac{1}{\sqrt{2}}|\tE_{\v k}|(|\tE_{\v k}^2-\Delta_{\v k}^2|
-\tE_{\v k}^2+\Delta_{\v k}^2)^\frac{1}{2}}
{|\tE_{\v k}^2-\Delta_{\v k}^2|}\Bigr\ra_v
\label{HDOS}
\ee
where $\tE_{\v k}$ is given by Eq.~(\ref{ETIL}) and the vortex average $\la
..\ra_v$ is defined in Eq.~(\ref{VORAV}). Note that N(E,H,$\theta$) also
depends on the field strength H via the DS, this variable is sometimes
not written explicitly. From Eq.~(\ref{HDOS}) the field-angle and
temperature dependence of the specific heat C(T,H,$\theta$) may
be obtained in the usual way. Defining the linear specific heat
coefficient as $\gamma$(T) = C/T and using
$\gamma_n=\frac{\pi^2}{3}N_0$ we obtain after a variable substitution
($\epsilon = E/T$)
\be
\frac{C(T,H,\theta)}{C_n}=\frac{\gamma(T,H,\theta)}{\gamma_n}
=\frac{3}{2\pi^2N_0}
\int_0^\infty d\epsilon\frac{\epsilon^2}{\cosh^2(\epsilon/2)}
N(T\epsilon,H,\theta)
\label{SPEC}
\ee
Together Eqs.~(\ref{ETIL}),(\ref{HDOS}) and (\ref{SPEC}) allow to calculate the
field-angle dependence
of the specific heat in the vortex phase. These equations are valid
within the DS approximation for the superclean limit for any
anisotropic gap function \De.

Altogether a fivefold integration over momenta ($\tthe,\tphi$), vortex
coordinates ($\trho,\alpha_v$) and energy $\epsilon$ has to be
performed in general. For T $\rightarrow$ 0 one needs only the
ZEDOS N(0,H,$\theta$) and only four integrations are left.
At this stage we have to proceed with numerical calculations to give
definite quantitative predictions. Approximate
analytical evaluations usually give only angle dependences but not the
absolute magnitude of the DS effect.
%
\begin{figure}
\includegraphics[clip=true,width=70mm]{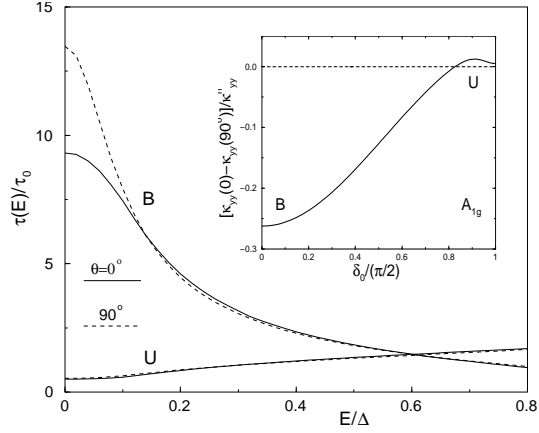}
\caption{Effective quasiparticle lifetime (Eq.~\ref{LIFE}) as function of
energy in Born approximation (B) and unitary (U) case for fields
parallel ($\theta$=0) and perpendicular ($\theta$=90$^\circ$) to
c-axis (H = 1T). The lifetime and its angular variation are much larger in the
Born case. As shown in the inset ($\delta_0$ = scattering phase
shift) this leads to a much larger thermal conductivity oscillation amplitude 
$[\hat{\kappa}_{yy}(0)-\hat{\kappa}_{yy}(90^\circ)]_{T=0}$ of up to
27\% in the Born(B) case, whereas it is only $\sim$ 1.2\% in the
unitary(U) case. The amplitude changes sign at
$\delta=0.83(\frac{\pi}{2})$ and is maximal at $\delta_0\simeq
0.91\frac{\pi}{2}$.}
\label{fig:tau}
\end{figure}
%
\subsection{Magnetothermal conductivity}
\label{subsect:KAPPA}

It is an important advantage of the DS method as compared to the more
advanced semiclassical methods \cite{Dahm02,Udagawa04} that it provides
expressions for both specific heat and thermal conductivity, whereas
the latter
method sofar can only be used for the DOS and specific heat. In the
zero-field case the thermal conductivity in the SC phase  can be
calculated within the linear response approach of Ambegaokar and
Griffin \cite{Ambegaokar65}. Using the DS approximation this has later
been extended to the vortex phase \cite{Barash97,Barash98,Won00}. 

In the normal state one has for the thermal conductivity
\be
\kappa^n_{ii}(T)=\frac{2\pi^2}{3}N_0\tau_0\la (v_i)^2\ra_{FS}T
\label{KAPPN}
\ee
where $\tau_0=\frac{\hbar}{2\Gamma}$ is the quasiparticle life time. In
the isotropic case $\la v_i^{2}\ra_{FS}$=$\frac{v_F^2}{3}$ and for
the anisotropic FS case the ratios 
$\la(v_z)^2\ra_{FS}/\la(v_\parallel)^2\ra_{FS}$ are given in
App.~\ref{app:QPVEL} by Eq.~(\ref{CVELA2}).

\begin{figure}
\includegraphics[clip=true,width=70mm]{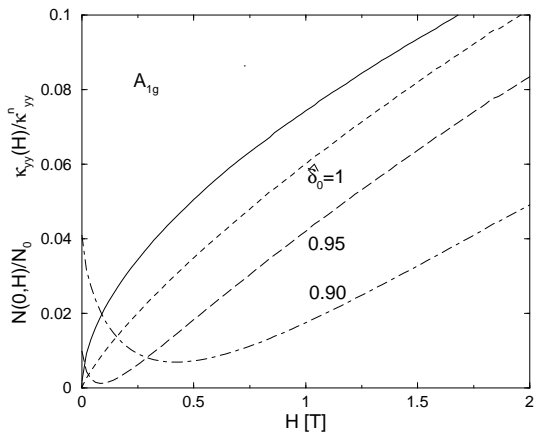}
\caption{
The A$_{1g}$ ZEDOS or specific heat $\gamma$(H,T=0)
coefficient as function of field strength for $\theta =90^\circ$ is
shown as full curve. Typical $\sqrt{H}$ of the low field ZEDOS is
observed. Thermal conductivity $\kappa_{yy}$(H,T=0) for $\theta =90^\circ$
normalised to normal state value $\kappa^n_{yy}$ is
shown for various scattering phase shifts
$\hat{\delta_0}=\delta/(\pi/2)$ close to the unitary limit
$\hat{\delta_0}$ = 1 (broken curves). Deviations from
$\hat{\delta_0}$ = 1 lead to an increasingly prominent nonmonotonic
behaviour of $\hat{\kappa}_{yy}(H)=\kappa_{yy}(H)/\kappa^n_{yy}$. The
nonmonotonic field dependence was observed in
\cite{Watanabe04}. For even smaller $\hat{\delta_0}$ approaching the
Born case the initial drop is preserved but then
$\hat{\kappa}_{yy}(H)$ becomes flat.}
\label{fig:baufield}
\end{figure}

In the presence of a SC gap the new energy scale introduced by
$\Delta$ leads to an energy dependent effective life time of
quasiparticles in the vortex state which is given by \cite{Mineevbook}
\be
\frac{\tau(E)}{\tau_0}\equiv\hat{\tau}(E)
&=&\frac{X(E)^2+Y(E)^2}{G_1(E)X(E)+G_2(E)Y(E)}\nn\\
X(E)&=&\cos^2\delta_0+\sin^2\delta_0
[G_1^2(E)-G_2^2(E)]\\
Y(E)&=&2\sin^2\delta_0G_1(E)G_2(E)\nn
\label{LIFE}
\ee
where G$_1$(E) and G$_2$(E) have been defined in
Eq.~(\ref{HDOS}). Furthermore $\delta_0$ is the (isotropic) scattering
phase shift which lies in the intervall [0,$\frac{\pi}{2}$]. In the
limiting cases one has:
\be
\delta_0=0&:&\qquad\hat{\tau}(E)=\frac{1}{G_1(E)}
\qquad\mbox{(Born limit)}\nn\\
\delta_0=\frac{\pi}{2}&:&\qquad\hat{\tau}(E)=\frac{G_1^2(E)+G_2^2(E)}{G_1(E)}
\qquad\mbox{(unitary limit)}
\label{BU}
\ee
The low energy behaviour of $\hat{\tau}$(E) in the Born and unitary
limit is quite
different as seen in Fig.~\ref{fig:tau}, therefore the low field
behaviour of the thermal conductivity is very sensitive to the size of
the scattering phase shift as discussed later (Fig.~\ref{fig:baufield}).

Thermal conductivity in the SC state for zero field involves
a quasiparticle energy integration and FS momentum averaging
\cite{Mineevbook}. This is
a special case of the magnetothermal conductivity in DS approximation
which we discuss here. The latter is obtained in the same spirit as the
field-angle dependent $\gamma(T,\v H)$-value: The SC quasiparticle
energies are replaced by their Doppler shifted values according to
Eq.~(\ref{ETIL}) and an additional averaging over the vortex coordinates
has to be performed. Then we obtain the final result ($\epsilon=E/T$)
\begin{eqnarray}
\hat{\kappa}_{ii}(T, \v H)=
\frac{\kappa_{ii}(T,\v H)}{\kappa^n_{ii}(T)} &=& \frac{3}{4\pi^2}
\int_0^\infty \frac{d\epsilon\epsilon^2}{\cosh^2(\epsilon/2)}
\hat{\tau}(T\epsilon)\frac{\langle\langle\hat{v}_{i\v k}^2
K(T\epsilon,\hat{\v k},\hat{\v r})\rangle\rangle_{FS,V}} 
{\langle\hat{v}_{i\v k}^2\rangle_{FS}}\nn\\
K(E,\hat{\v k},\hat{\v r}) &=& \frac{2}{|\hat{E}|}
[\hat{E}^2-\Delta(\hat{\v k})^2]^\frac{1}{2}
\Theta(\hat{E}^2-\Delta(\hat{\v k})^2)
\label{KAPPATH}
\end{eqnarray}
where $\Theta$ is the Heaviside function and
$\hat{E}=\Delta\tilde{E}_{\v k}(\theta,\trho,\alpha_v)$ whith
$\tilde{E}_{\v k}$ again given by Eq.~(\ref{ETIL}). Note that
i=$\tilde{x},\tilde{y},\tilde{z}$ refer to the fixed \emph{crystal}
coordinate system, although for brevity we will later use the conventional
notation $\kappa_{yy}$ etc. without the tilde. The above expression for
$\hat{\kappa}_{ii}$(T) is on the same level of approximation as
Eq.~(\ref{SPEC}) for $\gamma(T,\v H)$. It involves a five-fold integral
due the FS averaging ($\tthe,\tphi$), vortex ($v$) averaging
($\trho,\alpha_v$) and energy ($\epsilon$) integration. Finally we note
that the expression for the effective lifetime in Eq.~(\ref{LIFE})
is perturbative with respect to $\Gamma$. Since we consider only the
superclean limit where $\Delta|x|\gg\Gamma$ this is justified.\\

Due to the cylindrical symmetry of both FS (Fig.~\ref{fig:geometry})
and gap functions the calculated thermal conductivity depends only
on the relative angle $\phi$ between field rotational
($\tilde{x}\tilde{z}$) plane and heat
current \v j$_Q$. The $\phi$ dependence is caused by the factor
$\hat{v}^2_{i\v k}$ in the double average of Eq.~\ref{KAPPATH}
For \v j$_Q$ parallel ($\kappa_{xx}$) or perpendicular ($\kappa_{yy}$)
to the field rotation plane one
has $\hat{v}^2_{x\v k}=\hat{v}_a(k_z)^2\cos^2(\tphi)$ and 
$\hat{v}^2_{y\v k}=\hat{v}_a(k_z)^2\sin^2(\tphi)$ respectively. For
general angle $\phi$ one has to use $\hat{v}^2_{\phi\v k} =
\hat{v}_a(k_z)^2\cos^2(\tphi-\phi)$. Experimentally the perpendicular
configuration with $\phi=90^\circ$ has been used \cite{Watanabe04},
therefore we focus on $\kappa_{yy}(\theta,H)$.
%

\section{Application to \UPD}
\label{sect:UPDAL}

The DS-based calculation scheme for magnetothermal properties
described in detail before will now be applied to discuss recent
field-angle resolved thermal conductivity measurements in
\UPD~\cite{Watanabe04}. In this work it was established that the gap
function of \UPD~possesses a line node in the basal plane by a
qualitative discussion of the experimental results. It was also argued
that the experiment cannot
distinguish between several possible gap functions proposed in
\cite{McHale04} which have different positions of the node line along
the hexagonal axis. This was attributed to the \UPD~FS geometry which
is characterised by a dominating corrugated cylinder sheet oriented
along c$^*$. Later a further proposal implying a gap function with a
node line lying in a non-symmetry plane was made \cite{Won04a}. Before
discussing the results for these models of \De~we summarise their
basic symmetry properties and microscopic background.
%
\begin{table}
\caption{Spin and orbital structure of the possible gap functions
which are solutions of the \'{E}liashberg equations for the dual model
of UPd$_2$Al$_3$. The form factors of the anisotropic gap function
$\Delta(k_z)=\Delta\Phi(k_z)$ are assumed to have cylindrical symmetry. 
The state in the last
row \cite{Won04a} has a hybrid gap function since $\Phi(k_z)$ =
$2\cos^2c k_z$-1. The line nodes  $k_z=\pm\pi/4c$ are due to fine
tuning of the amplitudes of two fully summetric basis functions.
}
\begin{ruledtabular}
\begin{tabular}{lllllll}
$p$ & spin pairing & $|\chi\rangle = |S,\,S_z\rangle$ & D$_{6h}$
repres. & $\Phi(k_z)$ & nodal plane & type \protect\cite{Watanabe04}\\
\hline
-1 & OSP & $|1,0\rangle$ = $\frac{1}{\sqrt{2}} \left( |\!\uparrow\downarrow\rangle +
|\!\downarrow\uparrow\rangle \right)$
&A$_{1u}$  & $\sin(c k_z)$ & $k_z=0$ & I\\
-1 & OSP & $|0,0\rangle$ = $\frac{1}{\sqrt{2}} \left(
|\!\uparrow\downarrow\rangle -|\!\downarrow\uparrow\rangle \right)$ 
&A$_{1g}$ & $\cos(c k_z)$ & $k_z=\pm\frac{\pi}{2c}$ & II\\
+1 & ESP & $|1,\pm1\rangle$ = $|\!\uparrow\uparrow\rangle,
|\!\downarrow\downarrow\rangle$ 
&A'$_{1u}$ & $\sin(2c k_z)$ & $k_z=0,\pm\frac{\pi}{2c}$ & III\\ 
\hline
-1 & OSP& $|0,0\rangle$ = $\frac{1}{\sqrt{2}} \left(
|\!\uparrow\downarrow\rangle -|\!\downarrow\uparrow\rangle \right)$ 
& A$_{1g}\oplus$ A'$_{1g}$ &  $\cos(2c k_z)$ & $k_z=\pm\frac{\pi}{4c}$ & IV
\end{tabular}
\end{ruledtabular}
\label{tab:GAPFUN}
\end{table}
%
\begin{figure}
\includegraphics[clip=true,width=70mm]{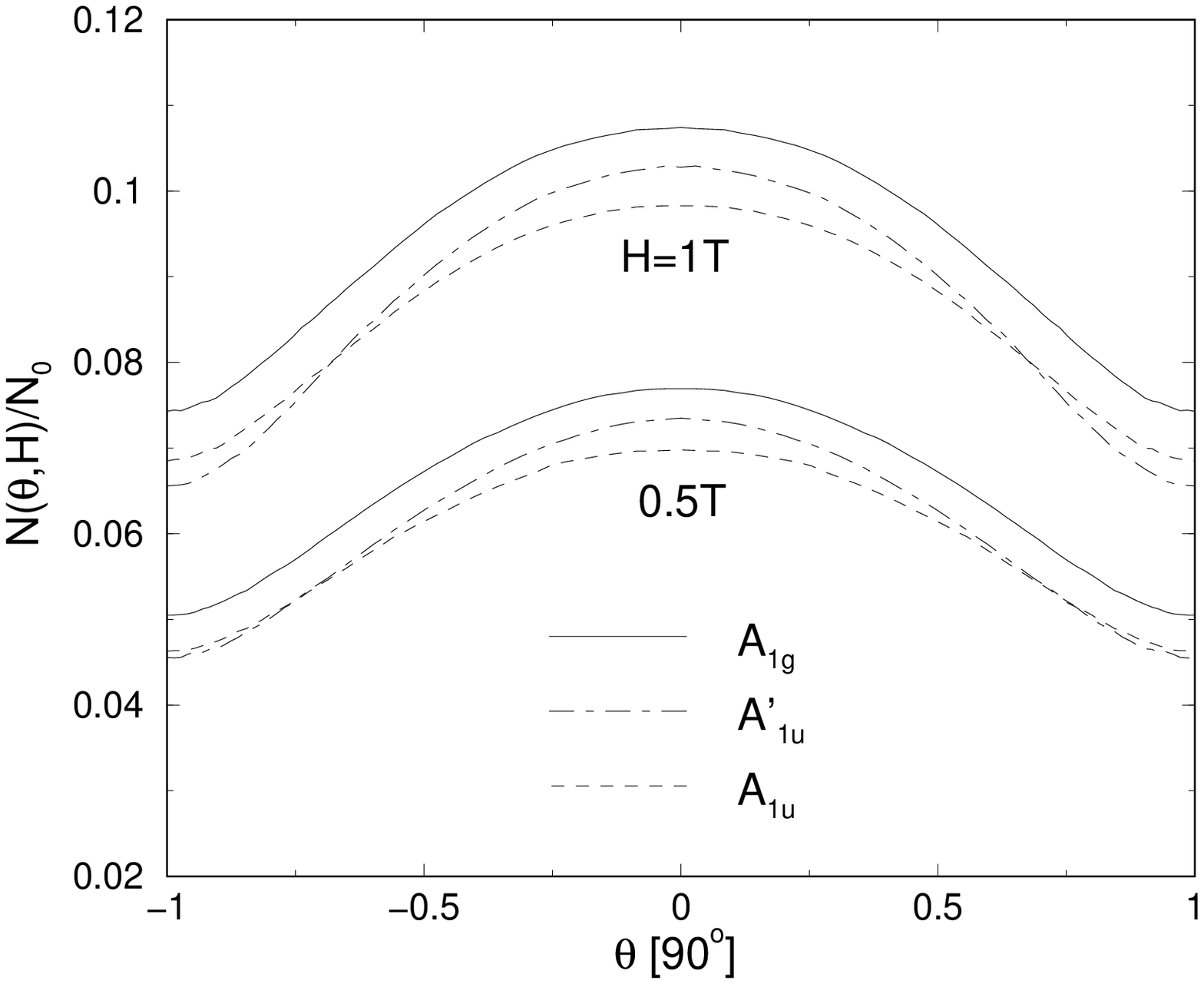}
\caption{Dependence of ZEDOS or specific heat $\gamma$(\v
H)- coefficient on vortex direction $\theta$ at different field
strengths. Qualitative behaviour
for all type I-III gap functions is equivalent, this also holds for
$\kappa_{yy}(\theta,H)$.}
\label{fig:densang}
\end{figure}
%
\subsection{The symmetry properties of gap function candidates}
\label{subsect:GAP}

\UPD~is the only HF superconductor whose microscopic mechanism of
Cooper pair formation is known with some certainty. As mentioned in
the introduction this is due to the partly itinerant and partly
localised nature of 5f-electrons. The latter lead to a well defined
magnetic exciton band seen in INS experiments. Most crucially the
magnetic exciton at the AF wave vector \v Q is also seen in a strong
coupling signature of the tunneling DOS of conduction electrons at an
energy ($\simeq$ 1 meV) that is identical to the INS results. This is
strong evidence that magnetic excitons originating in CEF excitations
of the local 5f subsystem mediate the Cooper pairing of itinerant 5f
electrons. This mechanism has been investigated both in weak coupling
\cite{Thalmeier02} and in a strong-coupling Eliashberg approach
\cite{McHale04}. In the latter a model for the effective interaction
based on magnetic exciton exchange with Ising type spin space symmetry
was proposed. This breaks rotational symmetry in the (pseudo-) spin
space of conduction electrons in a maximal way, therefore the pair
states $|\chi\ra$ have to be classified according to equal spin
pairing (ESP) and
opposite spin pairing (OSP) states characterised by a
spin projection factor p =$\la\uparrow\downarrow|
\sigma_z\sigma_z|\uparrow\downarrow\ra$ rather than in terms
of singlet and triplet pairs. It was found \cite{McHale04} that three
of these states (type I-III) have a finite T$_c$. The largest T$_c$
belongs to two
degenerate OSP states of opposite parity. These states together with
their orbital dependence and symmetry classification are tabulated
in Table~\ref{tab:GAPFUN}. In addition we have included a hybrid gap
function (last row) of type IV proposed in \cite{Won04a} consisting of a
superposition of two inequivalent fully symmetric D$_{6h}$
representations. Due to this fact its nodal lines are lying in
non-symmetry planes. This is rather uncommon feature and not observed
in any unconventional SC sofar. In addition this gap function does not appear as a
possible solution of the Eliashberg equations in the model of \cite{McHale04}.
Nevertheless we include it in the present discussion because it
was proposed as a candidate in \cite{Won04a}.

\subsection{Results of numerical calculations}
\label{subsect:NUM}

In the following we discuss the numerical results using the above
analysis for DOS, specific heat and thermal conductivity. We will use
the model parameters $\gamma_c$ = 0.8 for
the corrugated cylindrical FS and $\alpha$ = 0.69 for the anisotropy of
the Fermi velocity appropriate for \UPD. These parameters are taken as
independent but in principle they are related via Eq.~(\ref{CVELA1}). The
impurity scattering phase shift $\delta_0$ which is the only free or
unknown parameter in the theory is mostly taken to be close to its
unitary limit $\pi$/2 which is commonly assumed for HF compounds. We
choose a value $\delta_0=0.9(\pi/2)$ for which the
oscillation amplitude of $\kappa_{yy}(\theta,H)$ is close to its
maximum. However we also study the thermal conductivity for more
general $\delta_0$. Because we consider the superclean limit the ZEDOS is not
influenced by the choice of $\delta_0$. We will discuss the typical results
for the various gap functions in Table~\ref{tab:GAPFUN} but will not
present an exhaustive overview of the results. The intention is rather
to investigate whether the classification of gap functions introduced
in \cite{Watanabe04} is justified. The four gap function examples in
Table~\ref{tab:GAPFUN} correspond to the different types I-IV of nodal
structure whose field-angle dependence of specific heat and thermal
conductivity has been qualitatively discussed already in
\cite{Watanabe04}. Using the theory outlined in the previous sections
we can now perform detailed numerical calculations and check the
conjectures given in \cite{Watanabe04} quantitatively.\\

The important effect of the Doppler shift of quasiparticle energies
shown in Fig.~\ref{fig:dopptot} is the appearance of a non-vanishing
ZEDOS (Fig.~\ref{fig:dos}). Since the DS depends strongly on the
quasiparticle momentum \v k for a given field direction the ZEDOS, which
is dominated by quasiparticles in the nodal regions, will
exhibit pronounced field-angle dependence in addition to its dependence on
field strength. To simplify the discussions in the following we do not
distinguish any more between field ($\theta_H$) and vortex ($\theta$)
directions since they are very close for the present value of
$\alpha$ (see inset of Fig.~\ref{fig:dopptot}). The field dependence
of ZEDOS or specific heat $\gamma$-coefficient is
shown by the full curve in Fig.~\ref{fig:baufield} for the
A$_{1g}$ gap function at $\theta=90^\circ$. It exhibits the typical
$\sim\sqrt{H}$-behaviour for nodal gap functions which is due to the
DS shifted continuum states in the inter-vortex region as first
predicted by Volovik \cite{Volovik93}. This is in contrast to the
$\sim$ H behaviour of the specific heat coefficient in isotropic
superconductors for H $\ll$ H$_{c2}$ which is due to the quasi-bound
states in the vortex cores. The field
dependence of $\kappa_{yy}$ for $\theta$=90$^\circ$ is shown in the same
Fig.~\ref{fig:baufield} for various scattering phase shifts. The
behaviour is closer to linear H-dependence. If the scattering
phase shift deviates from the unitary limit just
a few percent, then immediately a nonmonotonic low-field behaviour
with an associated minimum in $\kappa_{yy}(H)$ appears as is obvious from
Fig.~\ref{fig:baufield}. It is caused by the low energy behaviour in
the effective quasiparticle life time shown in
Fig.~\ref{fig:tau}. Such nonmonotonic behaviour has indeed been found
experimentally in \cite{Watanabe04}.\\

\begin{figure}
\includegraphics[clip=true,width=70mm]{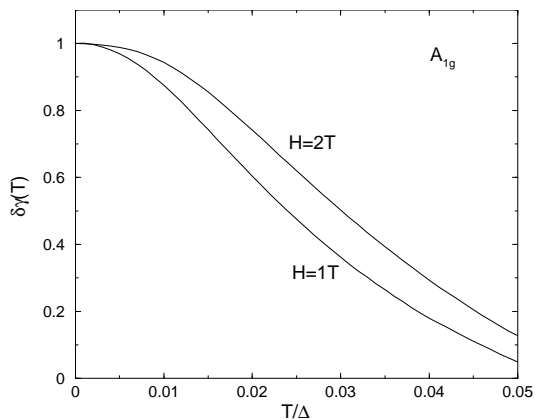}
\caption{Temperature dependence of the normalised field-angular
oscillation amplitude for specific heat or ZEDOS at different fields.
The amplitude is defined by $\delta\gamma=\delta
N=[N(0)-N(90^\circ)]_T/[N(0)-N(90^\circ)]_0$.}
\label{fig:temp}
\end{figure}

The field-angle dependence of the ZEDOS or specific heat $\gamma$-value
as shown in
Fig.~\ref{fig:densang} is rather directly determined by that of the
DS of nodal quasiparticles as may be seen by a comparison with
Fig.~\ref{fig:dopptot} keeping in mind the fact that it becomes
large for a field angle when quasiparticles with \emph{all} \v k -
vectors are Doppler shifted as is the case for $\theta$=0. The maximum
of the oscillation observed for this angle increases with $\sqrt{H}$
behaviour as mentioned before. For fields reasonably well below
H$_{c2}$ the oscillation amplitude is of the order of several per cent
of the normal state DOS N$_0$ or normal state $\gamma_n$
(Fig.~\ref{fig:densang}). In this figure we also show a comparison of angular
dependence for the first three order parameters (type I-III) in
Table~\ref{tab:GAPFUN} which all have node lines parallel to the hexagonal
plane but at different values of k$_z$. In Ref.~\cite{Watanabe04} it was
suggested with qualitative arguments that the angle dependences should
be similar in the three cases. The reason is that the Doppler shift is
determined by the product $\v v_s\cdot\v v_F$ and $\v v_F$ is always
parallel to the hexagonal plane for the possible node line positions $k_z$
= $\pm\frac{\pi}{2},0$ of type I-III. Since the latter are reflection
planes the Fermi velocity perpendicular to these planes has to vanish
as is obvious from
the corrugated cylinder FS of Fig.~\ref{fig:geometry}. This qualitative
expectation is indeed confirmed by the numerical results for type
I-III gap functions in Fig.~\ref{fig:densang} which shows the same
type of $\theta$ dependence. The small difference of the amplitude is
due to the different size of the Fermi velocities at the above node
line positions (Eq.~\ref{CQPVEL}) due to FS corrugation. The present
fully numerical
treatment of the DS theory allows also to calculate the temperature
dependence of the oscillation amplitude of $\gamma(\theta,T)$;
it is shown in Fig.~\ref{fig:temp}. 
The amplitude first decreases with T$^2$ behaviour and
then decreases rapidly. 
For \UPD~ one has 2$\Delta_{av}$/kT$_c$ = 5.6 \cite{Sato01},
therefore (T/T$_c$) = 2.8(T/$\Delta_{av}$). Here $\Delta_{av}$ should
be interpreted as a FS-sheet and momentum averaged gap value. If we only
consider the cylindrical FS sheet, then for all of our gap models
(I-III) we have  $\Delta_{av}$ =$\Delta$/$\sqrt{2}$ which means
(T/$\Delta$)=0.25(T/T$_c$). Comparison with Fig.~\ref{fig:temp} shows
that the T=0 oscillation amplitude has dropped to 10-15\% of its
original value when the temperature has increased to 20\% of
T$_c$. This illustrates the neccessity of having T$\ll$T$_c$ if one
wants to observe DS induced angular oscillations of magnetothermal quantities.\\
%
%

\begin{figure}
\includegraphics[clip=true,width=70mm]{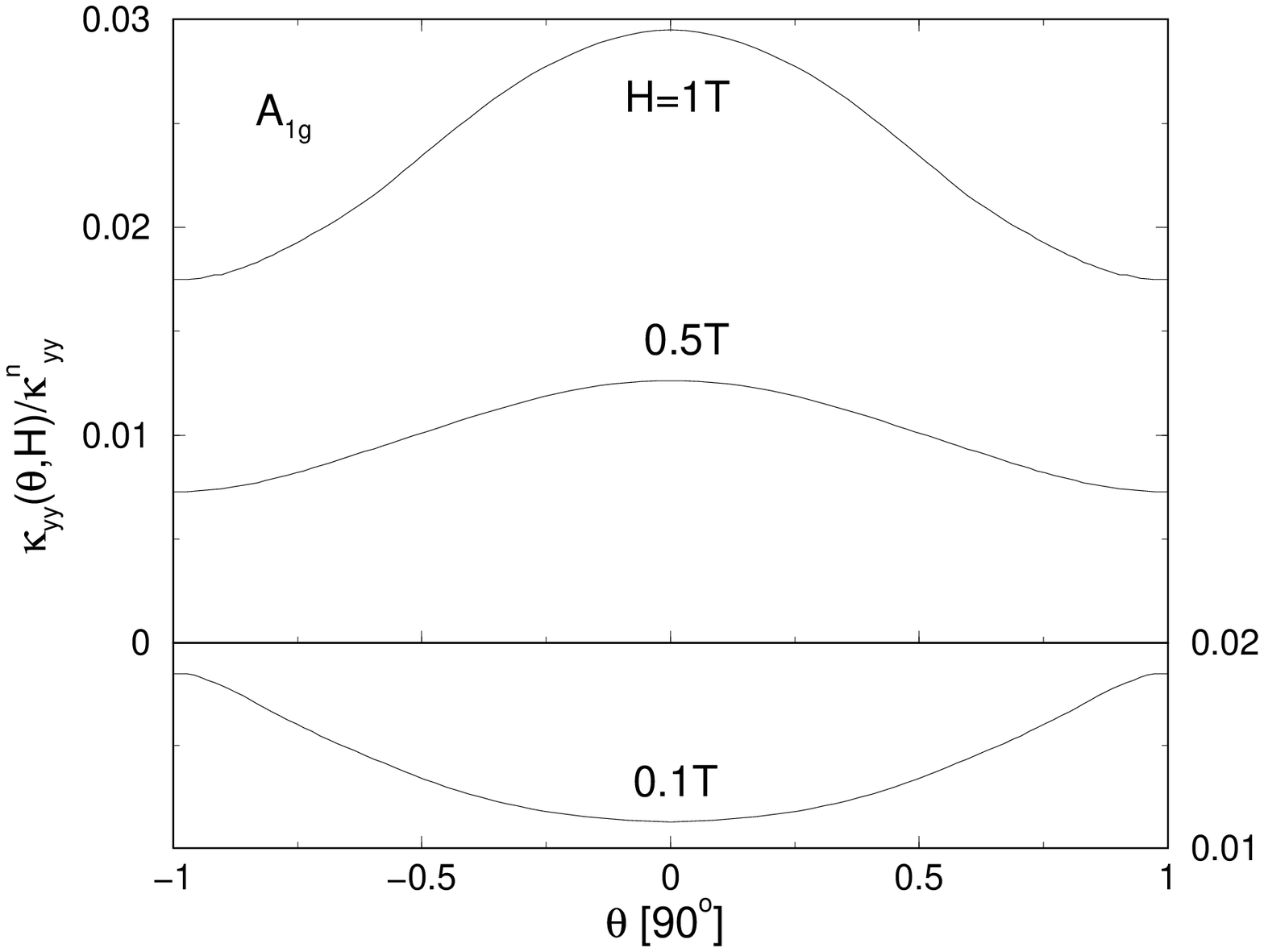}\hfill
\includegraphics[clip=true,width=70mm]{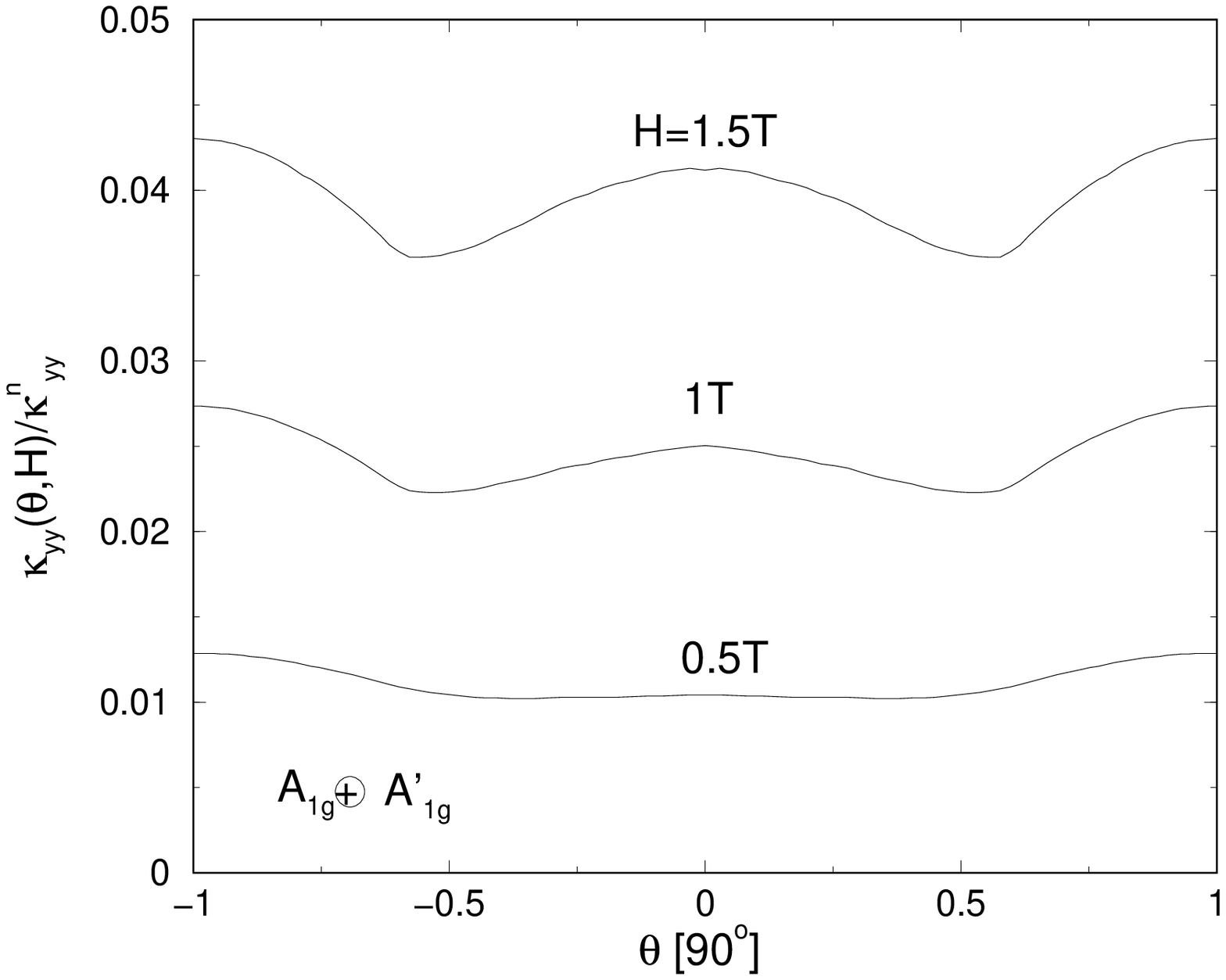}
\caption{Left panel: Thermal conductivity as function of field angle
for type II gap function at various field
strengths. This behaviour, notably the sign change of the oscillation
amplitude with increasing field  corresponds to experimental
observations for fields H $\leq$ 2T in \cite{Watanabe04}.
Type I,III gap functions (not shown here) exhibit very
similar oscillations as shown in the case of ZEDOS
(Fig.~\ref{fig:densang}). Right panel: Thermal conductivity for the
hybrid (type IV) gap
function \De = $\Delta\cos 2ck_z$. A sharp minimum at intermediate
$\theta\simeq 51^\circ$ due to off-symmetry nodal plane appears. This
behaviour is in contradiction to experimental observations \cite{Watanabe04}.}
\label{fig:coxxang}
\end{figure}

Similar observations can be made for the thermal conductivity
$\kappa_{ii}(\theta)$ for type I-III gap functions as shown in
Fig.~\ref{fig:coxxang} where we discuss $\kappa_{yy}(\theta)$. This
component corresponds to the experimental configuration with heat
current perpendicular to the plane of field rotation. The calculation
has been done for the nearly unitary phase shift
$\delta_0$=0.9($\pi/2$) where the oscillation amplitude is
close to its (positive) maximum as shown in the inset of
Fig.~\ref{fig:tau}. For this $\delta$ value  $\kappa_{yy}(90^\circ,H)$
has a minimum around H$_m\sim$ 0.4T (Fig.~\ref{fig:baufield}). 
For fields above H$_m$ again the maximum appears at
$\theta$=0 when the field is perpendicular to the nodal plane.
However for fields below H$_m$ a sign change of the amplitude takes
place and therefore the maximum appears for field direction lying
within the nodal plane. This is due to the increased influence of life
time angular
dependence at small fields. The shape of angular oscillations, their
field sequence (Fig.~\ref{fig:baufield}, left panel) and the minimum
of $\kappa_{yy}(90^\circ,H)$ in Fig.~\ref{fig:baufield} correspond well to the
experimental observations in \cite{Watanabe04}.

The relative differences in the oscillation amplitude for type I-III
gap functions caused by FS corrugation are somewhat larger as in the
case of ZEDOS, however their qualitative behaviour is again
indistinguishable and is not shown here. The absolute oscillation
amplitude of the order of per cent of the normal state conductivity at
1T is smaller than for ZEDOS (Fig.~\ref{fig:densang}). In the present
superclean limit the
latter is independent of the scattering phase shift, i.e. unitary or
Born limit. In contrast the absolute oscillation amplitute of
$\hat{\kappa}(\theta,H)$ depends strongly on the phase shift $\delta_0$
via the pronounced $\delta_0$ dependence of the effective life time as
seen in Fig.~\ref{fig:tau}. As the inset of  Fig.~\ref{fig:tau}
shows, the oscillation amplitude for $\kappa_{yy}(\theta,H)$ is very
large (with opposite sign) in the Born limit and comparatively small
(of order per cent) in the (nearly) unitary limit which we have assumed here
for \UPD. A similar calculation for $\kappa_{xx}(\theta,H)$, i.e. heat
current parallel to the field rotation plane shows that the
oscillation amplitude is always positve and again much larger for Born
as compared to nearly unitary scattering. The calculated absolute magnitude of the
T=0 thermal conductivity  $\kappa_{yy}(\theta,H)$ is smaller by a
factor of three compared to the
experimental value at the lowest temperature measured. In addition a
similar T-dependendence as for ZEDOS (Fig.~\ref{fig:temp}) would lead
to a further strong reduction. These discrepancies may be linked
with our insufficient model for the energy and field dependence of the
quasiparticle life time. They cannot presently be resolved without 
experimental results on the T-dependence of the oscillation amplitudes.  
For even larger larger fields (H $\geq$ 2.5T) than in Fig.~\ref{fig:coxxang} an
additional sign change in the oscillation was observed. In this regime
the DS approximation breaks down due to vortex overlap and the sign
change was rather attributed to the influence of
the uniaxial H$_{c2}$-anisotropy \cite{Watanabe04}. 

Since angular oscillations are not observed when the field is rotated
in the hexagonal plane \cite{Watanabe04} (and heat current parallel to
c-axis) one may conclude that the node line of the gap is indeed parallel to the
hexagonal plane as for all type I-IV gap functions discussed
here. Our calculations prove quantitatively that the
observed  $\kappa_{yy}(\theta)$-dependence cannot distinguish between
the type I-III cases since the $\theta$ dependence for these gap
functions is very similar as has already been suggested in
\cite{Watanabe04} on qualitative grounds. The reason has been
discussed above in the context of ZEDOS (Fig.~\ref{fig:densang}) oscillations.\\
%
%

The situation however is different for the hybrid
A$_{1g}\otimes$A'$_{1g}$ gap function (type IV) which has node lines
in non-symmetry planes. The corresponding angle dependence of the thermal
conductivity $\kappa_{xx}(\theta)$ is shown in the right panel of
Fig.~\ref{fig:coxxang}. It has a completely different appearance from
those for the type I-III gap functions. Firstly, they do not drop to a
small value for field angle $\theta=\pm90^\circ$, instead they
are even larger than for $\theta$=0. Secondly a
pronounced  and sharp minimum appears at an intermediate angle
$\theta\simeq 51^\circ$. This is due to the
existence of a non-symmetry nodal plane for the
A$_{1g}\otimes$A'$_{1g}$ order parameter. In such case both
v$_{x,y}$(k$_z$) and v$_z$(k$_z$) are nonzero and they give contributions
to the DS in Eq.~(\ref{DOPPDEF}) which decrease or increase as function
of $\theta$ respectively, leading to the minimum at an intermediate
value that survives in the averaged quantities like thermal
conductivity. The minimum cannot be discussed away by including
the contribution from other FS parts. Firstly it is known from LDA calculations
\cite{Zwicknagl03} that the cylinder FS gives one of the dominant
contributions to the
total DOS, secondly the other, e.g. ellipisoidal sheets also have a finite
v$_z$ at a non-symmetry nodal plane and therefore would give a similar
behaviour. Consequently, if the gap function is of the type IV the
kink-like minimum at intermediate $\theta$ has to be present
in $\kappa_{xx}(\theta)$. 
However, as mentioned above, the experiments
\cite{Watanabe04} show that it behaves very much as expected for the
type I-III gap functions discussed before (Fig.~\ref{fig:coxxang}, left
panel). No trace of a minimum or only depression at intermediate
angles has been found. Therefore one
has to conclude that the experiments in \cite{Watanabe04} rule out the
type IV gap function \De=$\Delta\cos(2ck_z$) proposed in
\cite{Won04a} for \UPD. It was already suggested before that in any case
this is an unlikely candidate because it is a hybrid gap function with
nodes at non-symmetry planes, which has never been found in any other
unconventional superconductor.

\section{Summary and Outlook} 
\label{sect:OUTLOOK}

In this work we have investigated the ZEDOS, specific heat and
magnetothermal transport properties of superconducting \UPD~in the
vortex phase using the Doppler shift approximation. Our intention was
to give a quantitative basis to the qualitative discussion of possible
gap function symmetries presented together with the experimental
results of \cite{Watanabe04}. We have
given a coherent presentation of the known Doppler shift analysis and
expressions and performed the evaluation of physical quantities in a
fully numerical approach which allows quantitative predictions. The DS
approach is oversimplified in the sense that it does not correctly
account for the vortex core contributions and the effect of vortex
overlap on approaching H$_{c2}$. Therefore one is limited to
fields reasonably well below H$_{c2}$ but still large enough for the
superclean limit to hold. On the other hand it has the
great advantage as compared to more advanced semiclassical methods
that transport quantities and not only the quasiparticle DOS are
easily accessible within a linear response treatment.
We have made predictions for the candidate type I-IV gap functions of
\UPD~which have all line nodes parallel to the hexagonal plane but
with different multiplicity and position along k$_z$. Our quantitative
analysis fully confirms the qualitative conjectures drawn in
\cite{Watanabe04}. Only gap functions which have nodal lines in
symmetry planes containing the BZ boundary or center are compatible with
experimental results, those with
nodal lines in off-symmetry planes can clearly be ruled out. 
It was suggested that the nonmonotonic behaviour of
the low-field thermal conductivity may be caused by the sensitivity of
the effective life time to deviations from unitary scattering. This is
also connected with the sign change of the oscillation amplitude for
small fields.\\

The angular-resolved thermal
conductivity thus proves that a node line of \De~must be present in a
hexagonal symmetry plane, however by itself it cannot distinguish
between the possible gap functions of type I-III. As explained in
\cite{Watanabe04} one needs additional information: Inelastic neutron
scattering \cite{Bernhoeft00} requires the translation symmetry
$\Delta(\v k +\v Q)$ = -\De~which would exclude A'$_{1u}$, also this
gap function is disfavored in Eliashberg theory
\cite{McHale04}. Furthermore naive interpretation of Knight shift
results \cite{Tou95} advocates for an even parity gap function,
although a comparitive analysis for the A$_{1g}$ and A$_{1u}$ states
has not been performed \cite{McHale04} yet. This finally led to the
suggestion \cite{Watanabe04} that the A$_{1g}$ gap function in the
second row of Table~\ref{tab:GAPFUN} is the proper one for \UPD.\\ 

Presently we have restricted ourselves to gap functions
$\Delta(\tthe)$ that have cylindrical symmetry. Our numerical
method can straightforwardly be applied to fully anisotropic gap
functions $\Delta(\tthe,\tphi)$. The calculation then provides us with
a direct mapping of $\Delta(\tthe,\tphi)\rightarrow \gamma(\theta_H,\phi_H)$
or $\kappa_{ii}(\theta_H,\phi_H)$ between the momentum dependence of gap
functions and the field-angle dependence of physical quantities in
their respective 2D domains. Of course this mapping has no unique inverse,
but nevertheless comparison with the experimental $\theta_H,\phi_H$-
dependence may provide important clues on the anisotropy character of
the gap functions. In addition, an extension of the present quantitative
theory to different types of FS sheets like FS ellipsoids,
corrugated tight binding with FS nesting features etc. is easily possible.  
Finally we note that the numerical approach to the
DS theory may in principle be generalised beyond the perturbative
treatment of scattering, i.e. from the superclean to the clean
limit with $\Delta |x| <\Gamma \ll \Delta$, although this would likely
mean a much larger computational effort.


\appendix
\section{Quasiparticle velocities for corrugated cylindrical FS}
\label{app:QPVEL}

In this appendix we define the geometric features and quasiparticle
properties of the corrugated cylindrical Fermi surface
(FS) which is necessary for the calculation of the Doppler shift
energies. Specifically we give the quasiparticle velocities in terms
cylindrical coordinates. The corrugated cylinder is the most prominent
heavy FS sheet in \UPD~obtained in LDA calculations
\cite{Knoepfle96,Inada99}, dual model calculations \cite{Zwicknagl03}
and also from dHvA experiments \cite{Inada99}. The latter show it has
also among the heaviest quasiparticle masses. In the AF BZ
($-\frac{\pi}{\tilde{c}}\leq k_z\leq\frac{\pi}{\tilde{c}}$) with
$\tilde{c}$ = 2c appropriate for \UPD~it can be modeled as
\begin{eqnarray}
\epsilon(\v k)=\epsilon_a(\v k) - 2 t_c\cos\tilde{c}k_z
\qquad,\qquad
\epsilon_a(\v k)=\frac{\hbar^2}{2m_a}(k_x^2+k_y^2) 
\label{CDISP}
\end{eqnarray} 
where the first part is the parabolic ab-plane dispersion and the
second part is the tight-binding like dispersion which determines the FS
corrugation along c. The diameter of the corrugated cylinder is given
by
\begin{eqnarray}
k_F^a(k_z)=k_F^0[1+\frac{2t_c}{\epsilon_F}\cos(\tilde{c}k_z)]^\frac{1}{2}
\qquad;\qquad
k_F^0=\bigl(\frac{2m_a\epsilon_F}{\hbar}\bigr)^\frac{1}{2}
\label{CFERMI}
\end{eqnarray}
We introduce the FS corrugation parameter by
\begin{eqnarray}
\gamma_c=\frac{k_F^a(0)}{k_F^a(\frac{\pi}{\tilde{c}})}
\qquad;\qquad
\lambda=\frac{2t_c}{\epsilon_F}
\label{CCORR1}
\end{eqnarray}
where $\gamma_c$ is the ratio of FS cross sectional areas at the (AF)
zone center and zone boundary respectively. It is given by
\begin{eqnarray}
\gamma_c=\frac{1+\lambda}{1-\lambda}
\qquad \mbox{or} \qquad 
\lambda = \frac{\gamma_c-1}{\gamma_c+1}
\label{CCORR2}
\end{eqnarray}
The quasiparticle velocities \v v$_{\v k}$ =
$\frac{1}{\hbar}\nabla\epsilon_{\v k}$ in cylindrical coordinates are
given by
\begin{eqnarray}
v_x(k_z,\tphi)&=&v_a^0\hv_a(k_z)\cos\tphi\nn\\
v_y(k_z,\tphi)&=&v_a^0\hv_a(k_z)\sin\tphi\nn\\
v_z(k_z)&=&v_c^0\hv_c(k_z)\nn\\
v_a^0&=&\hbar k_F^0/m_a\\
v_c^0&=&2\pi t_c/\hbar\tQ\nn\\
\hv_a(k_z)&=&[1+\lambda\cos\tilde{c}k_z]^\frac{1}{2}\nn\\
\hv_c(k_z)&=&\sin\tilde{c}k_z\nn
\label{CQPVEL}
\end{eqnarray}
the anisotropy ratio of the Fermi velocity of quasiparticles is then given by ($\tQ
\equiv\frac{\pi}{\tc}$): 
\begin{eqnarray}
\alpha=\bigl(\frac{v_c^0}{v_a^0}\bigr)^2=
\frac{\pi^2}{4}\bigl(\frac{k_F^0}{\tQ}\bigr)^2\lambda^2
\qquad \mbox{or} \qquad
\alpha=\frac{\pi^2}{4}\bigl(\frac{k_F^0}{\tQ}\bigr)^2
\bigl(\frac{\gamma_c-1}{\gamma_c+1}\bigr)^2
\label{CVELA1}
\end{eqnarray}
The anisotropy ratio $\alpha$ and corrugation factor $\gamma_c$ are
the parameters which determine the Doppler shift energy for the
present FS. The relation between $\alpha$ and $\gamma_c$ in
Eq.~(\ref{CVELA1}) is valid only for parabolic in-plane dispersion. It
is better to assume these parameters as independent and take their values
from experiment, keeping in mind that $\alpha\rightarrow$ 0 for
$\gamma_c\rightarrow$ 1.

Finally the FS averages over quasiparticle velocities are given by
\begin{eqnarray}
\la v^2_\parallel\ra_{FS}&=&\la v_x^2+v_y^2\ra_{FS}=(v_a^0)^2\nn\\
\la v_z^2\ra_{FS}&=&\frac{1}{2}(v_c^0)^2 \\
\frac{\la v_z^2\ra_{FS}}{\la
v^2_\parallel\ra_{FS}}&=&\frac{1}{2}\alpha \nn
\label{CVELA2}
\end{eqnarray}

\section{Critical fields, coherence length penetration depth in
uniaxial superconductors}
\label{app:HCRIT}

The theory of critical fields H$_{c1}$ and H$_{c2}$ in superconductors
with uniaxial symmetry like D$_{4h}$ and D$_{6h}$ was given in
\cite{Balatskii86,Bulaevskii90} (see also \cite{Abrikosovbook}) on the
basis of Ginzburg-Landau theory for
a single component SC order parameter. Here we give a summary of
relations derived in these references which are important in our
context of Doppler shift calculations. In uniaxial geometry shown in
Fig.~\ref{fig:geometry} coherence length and penetration depth are different
for fields directed along a or c crystal axes. As a consequence, for
intermediate polar field angle $\theta_H$ (with respect to c) in the
range $0\leq\theta_H\leq\frac{\pi}{2}$ the field ($\theta_H$) and
vortex ($\theta$) directions are not the same. This is an essential
difference to the isotropic case where field and vortices are
alligned. For uniaxial effective masses m$_a$, m$_c$ these angles are
related by 
\begin{eqnarray}
\theta=\tan^{-1}(\frac{1}{\alpha}\tan\theta_H)
\quad;\quad
\theta_H=\tan^{-1}(\alpha\tan\theta)
\quad;\quad
\alpha=\frac{m_a}{m_c}=\bigl(\frac{v_c^0}{v_a^0}\bigr)^2
\label{ANGTRANS}
\end{eqnarray}
where we assume the convention
$-\frac{\pi}{2}\leq\theta,\theta_H\leq\frac{\pi}{2}$ for the vortex
and field direction. The function $\theta(\theta_H)$ is plotted in the
inset of 
Fig.~\ref{fig:dopptot} for the effective mass anisotropy $\alpha = 0.69$
which is the appropriate average value for \UPD. For $\alpha$-values
only moderately different from one, $\theta$ and $\theta_H$ are rather
close. However, for $\alpha\ll 1$ the vortices
($\theta$) are 'pinned' along the c-axis and they are lagging behind
the field direction ($\theta_H$) when it is continuously changed from
c to a. The direction dependence of critical fields is given by
\begin{eqnarray}
H_{c2}(\theta_H)&=&\frac{\Phi_0}{2\pi\xi^2}
\quad;\quad
\Phi_0=\frac{hc}{2e}\nn\\
H_{c1}(\theta_H)&=&\Bigl[\frac{H^a_{c1}H^c_{c1}}
{[(H^a_{c1})^2\cos\theta +(H^c_{c1})^2\sin\theta]}\Bigr]^\frac{1}{2}
\quad;\quad \theta = \theta(\theta_H)
\label{HCRIT}
\end{eqnarray}
Where $\xi$ is given in Eq.~\ref{COH2} and the uniaxial H$_{c1}^{a,c}$ 
are obtained from 
\be
H^c_{c1}=\frac{\Phi_0}{2\pi}\frac{\ln\kappa_a}{\lambda_a^2}
\quad;\quad
H^a_{c1}=\frac{\Phi_0}{2\pi}
\frac{\ln(\kappa_a\kappa_c)^\frac{1}{2}}{\lambda_a\lambda_c}
\quad \mbox{with} \quad
\kappa_{a,c}=\frac{\lambda_{a,c}}{\xi_{a,c}}
\label{HCRIT1}
\ee
The uniaxial Ginzburg-Landau coherence lengths $\xi_{a,c}$ and penetration depths
$\lambda_{a,c}$ are given by (i = a,c)
\be
\xi_i^2 = \frac{\hbar^2\la v_i^2\ra_{FS}}{(kT_c)^2|\tau|}
\quad \mbox{and} \quad
\lambda_i^2 = \frac{m_ic^2}{8\pi n_s(\tau)e^2}
\label{COH1}
\ee
with $\tau$=1-T/T$_c$. The FS averages for the corrugated cylindrical
FS are derived in App.~\ref{app:QPVEL}. The effective field-angle
($\theta_H$) dependent coherence length $\xi$ in
Eq.~(\ref{HCRIT}), penetration depth $\lambda$ and mass m are given by
($\lambda_\parallel\equiv\lambda_a, \xi_\parallel\equiv\xi_a,
m_\parallel\equiv m_a$)
\be
\xi=(\xi_\parallel\xi_\theta)^\frac{1}{2}
,\quad
\lambda=(\lambda_\parallel\lambda_\theta)^\frac{1}{2}
,\quad
m=(m_\parallel m_\theta)^\frac{1}{2}
\label{COH2}
\ee
where we used the definitions
\be
\xi^2_\theta&=&\xi_a^2\cos^2\theta + \xi_c^2\sin^2\theta\nn\\
\lambda^2_\theta&=&\lambda_a^2\cos^2\theta + \lambda_c^2\sin^2\theta
=\frac{m_\theta c^2}{8\pi n_se^2}\\
m_\theta&=&m_a\cos^2\theta +m_c\sin^2\theta\nn\\
\label{COHPEN}
\ee
Note that $\theta$ is the vortex angle with respect to the c-axis
which is related to the field angle $\theta_H$ via Eq.~(\ref{ANGTRANS}). From
the above equations we derive the relations
\be
\frac{\lambda_\parallel}{\lambda}&=&
\bigl(\frac{\lambda_\parallel}{\lambda_\theta}\bigr)^\frac{1}{2}=
\alpha^\frac{1}{4}[\sin^2\theta +\alpha\cos^2\theta]^{-\frac{1}{4}}\nn\\
\frac{\lambda_\theta}{\lambda}&=&
\bigl(\frac{\lambda_\theta}{\lambda_\parallel}\bigr)^\frac{1}{2}=
\alpha^{-\frac{1}{4}}[\sin^2\theta +\alpha\cos^2\theta]^{\frac{1}{4}}
\label{PEN}
\ee
which enter directly the expressions for the Doppler shift energies in
Eq.~(\ref{DOPPEXP}).

\section*{Acknowledgments}
The authors  would like to thank Kazumi Maki, Hyekyung Won and
Alexander Yaresko for collaboration.


\bibliography{ref.bib}

\end{document}